\documentclass[reprint,prd,onecolumn,amsmath,amssymb,12pt]{revtex4-2}   
\usepackage[utf8]{inputenc}
\usepackage{mwe,float}
\usepackage{caption,subcaption, url}
\usepackage{natbib}
\usepackage[nointegrals]{wasysym}
\usepackage{rotating}
\usepackage{array,tabu,multirow}
\usepackage{graphicx,bm}
\usepackage{dcolumn} 
\usepackage{floatpag}  
\rotfloatpagestyle{empty}
\usepackage{amsmath}
\usepackage{amsthm}
\usepackage{amssymb}
\usepackage{latexsym,upgreek}
\usepackage{hyperref,rotating}
\usepackage{xpatch,bigints}
\usepackage{booktabs,subcaption}
\usepackage{ulem}
\usepackage{textgreek}
\usepackage{mathrsfs}
\makeatletter
\AtBeginDocument{\xpatchcmd{\@thm}{\thm@headpunct{.}}{\thm@headpunct{\enskip}}{}{}}
\makeatother 
\usepackage{gensymb}
\usepackage{mathdots}

\usepackage[version=4]{mhchem}

\usepackage{graphicx,kecpm,slashed}
\usepackage{makeidx}
\makeindex
\usepackage{multirow}
\usepackage[dvipsnames]{xcolor}

\newcommand{\bex}{\begin{example}}
\newcommand{\eex}{\end{example}}
\newcommand{\besp}{\begin{split}}
\newcommand{\ensp}{\end{split}}

\newcommand{\by}{\times}

\newcommand{\bos}{\boldsymbol}

\newcommand{\btab}{\begin{tabular}}
\newcommand{\etab}{\end{tabular}}
\newcommand{\barr}{\begin{array}}
\newcommand{\earr}{\end{array}}
\newcommand{\bpm}{\begin{pmatrix}}
\newcommand{\epm}{\end{pmatrix}}
\newcommand{\bit}{\begin{itemize}}
\newcommand{\eit}{\end{itemize}}
\newcommand{\ben}{\begin{enumerate}}
\newcommand{\een}{\end{enumerate}}
\newcommand{\bct}{\begin{center}}
\newcommand{\ect}{\end{center}}

\newcommand{\bes}{\begin{split}}
\newcommand{\ens}{\end{split}}

\newcommand{\edoc}{\end{document}}

\newcommand{\tcdb}[1]{\textcolor{Blue}{#1}}

\newcommand{\tcfg}[1]{\textcolor{ForestGreen}{#1}}

\begin{document}
\title{A DESI universe with time-dependent dark energy is 119 Myr younger than a $\mathbf{\Lambda}$CDM universe}

\author{Kevin Cahill}
\email{cahill@unm.edu}

\affiliation{Department of Physics and Astronomy\\
University of New Mexico,
Albuquerque, New Mexico 87106}

\date{\today}

\nopagebreak

\begin{abstract}
On the basis of their redshift and Lyman-$\alpha$ measurements, the Dark Energy Spectroscopic Instrument (DESI) collaboration suggest that dark energy varies with the scale factor $a$ as $ \rho_{de}(a)={} \, a^{-3(1 + w_0 + w_a)} e^{-3 w_a (1-a)}\rho_{de,0} $ in which $w_0 = - 0.752 \pm 0.057$ and $w_a ={} - 0.86^{+0.23}_{-0.20} $.  Such a DESI universe is 119 Myr younger than the $\Lambda$CDM universe.
\end{abstract}

\maketitle 

\section{Time-dependent dark energy}
\label{Time-dependent dark energy}

In $\Lambda$CDM cosmology~\citep{AAplanck2024}, 
the mass density $\rho_{de}$
of dark energy 
is a constant multiple 
$ \Omega_\Lambda = 0.689$ 
\begin{equation}
\rho_{de} = \Omega_\Lambda \, \rho_c 
= 5.95  \by 10^{-27}~\text{kg m}^{-3}
\end{equation} 
of the critical density 
\begin{equation}
\rho_{c} ={} 3 H_0^2/(8\pi G) 
\simeq 
9.2 \by 10^{-27} \,
\text{kg\,m}^{-3} .
\label{critical density}
\end{equation} 
and is independent of 
the scale factor.
 Yet there is no reason to suppose that
 the density of dark energy is a constant 
 other than that constants 
 are invariant under 
 general coordinate transformations.
\par
The Dark Energy Spectroscopic Instrument (DESI) collaboration 
have measured the redshifts
of over 30 million galaxies and quasars 
and the Lyman-$\alpha$ forest spectra of more than 820,000 quasars~\citep{desicollaboration2025desidr2resultsii}. 
Combining their observations 
with the last Planck release  
(PR4) of CMB data~\citep{AAplanck2024}
and with the Pantheon+~\citep{Scolnic_2022},
Union3~\citep{rubin2024unionunitycosmology2000}, 
and ESY5~\citep{descollaboration2024darkenergysurveycosmology} surveys
of Type Ia 
supernovas,
they suggest that
the density of dark energy 
varies with the scale factor $a$ 
as
\begin{equation}
\frac{\rho_{de}(a)}{\rho_{de,0}}
\, ={} \, a^{-3(1 + w_0 + w_a)} \,
e^{-3 w_a (1-a)}  
\label {DESI's density}
\end{equation} 
in which~\citep{desicollaboration2025desidr2resultsii}
\begin{align}
w_0 ={}& - 0.838 \pm 0.055
\quad\text{and}\quad 
w_a = - 0.62 {}^{+0.22}_{-0.19}
\qquad \text{DESI+CMB+Pantheon+}
\label{Pantheon+}
\\
w_0 ={}& - 0.667 \pm 0.088
\quad\text{and}\quad 
w_a = - 1.09 {}^{+0.31}_{-0.27}
\qquad \text{DESI+CMB+Union3}
\label{Union3}\\
w_0 ={}& - 0.752 \pm 0.057
\quad\text{and}\quad 
w_a = - 0.86 {}^{+0.23}_{-0.20} 
\qquad \text{DESI+CMB+DESY5} .
\label{DESY5}
\end{align} 
The significance of their rejection 
of the ΛCDM parameters 
($w_0 = -1$ and $w_a = 0$)
is 2.8σ, 3.8σ, and 4.2σ
for combinations of their data 
with the Pantheon+, Union3, and DESY5
supernova samples respectively (Fig.~11 
of~\citep{desicollaboration2025desidr2resultsii}). 
\par
\begin{figure}[h!] 
\begin{center}
\textbf{Four dark-energy densities}\\
\includegraphics[width=5.0in, trim={0.in 0.in 0in 0.0in}, clip]{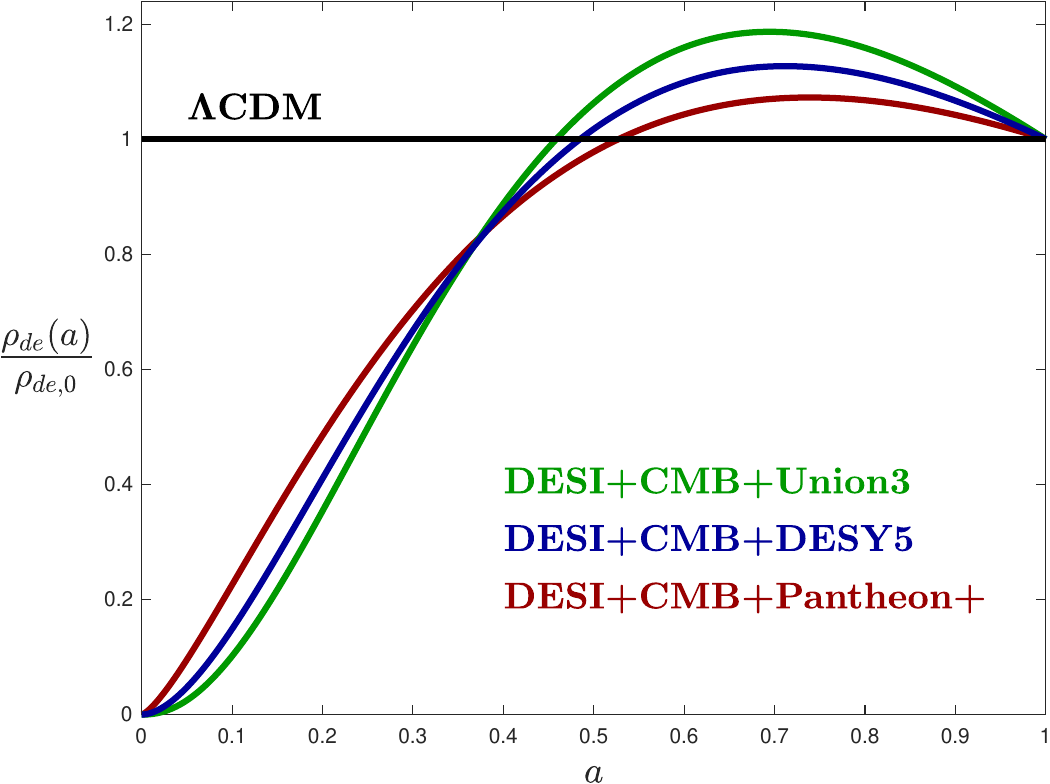}
\end{center}
\caption{
The ratio (\ref{DESI's density})  
of the DESI 
dark-energy density $\rho_{de}(a)$ 
to its present value $\rho_{de,0}$ 
is plotted against the scale factor $a$
for the $w_0, w_a$ values
\textcolor{RawSienna}{DESI+CMB+Pantheon+}
(\ref{Pantheon+}),
\tcfg{DESI+CMB+Union3} (\ref{Union3}), and
\tcdb{DESI+CMB+DESY5} (\ref{DESY5}).
Also shown is the constant
ratio 
$\rho_\Lambda/\rho_{de,0} = 1$  
of $\Lambda$CDM cosmology.}
\label {fig1}
\end{figure}

The ratio (\ref{DESI's density})  
of the DESI 
dark-energy density $\rho_{de}(a)$ 
to its present value $\rho_{de,0}$ 
is plotted against the scale factor $a$ 
in Fig.~\ref{fig1}. 

\section{Four cosmologies}
\label{Friedmann-Lemaitre-Robinson-Walker cosmologies}
\par
The Hubble frequency $H(a)$ varies 
with the scale factor $a$ as 
\begin{equation}
H(a) ={} \sqrt{\frac{8 \pi G}{3} 
\Big( \rho_\Lambda(a) + \rho_K(a) + \rho_m{a} + \rho_r(a) \Big)} .
\label {first-order Friedmann equation}
\end{equation}
Its present value,
the Hubble constant
$H_0 = H(a_0) = H(1)$, 
is approximately 
\begin{equation}
H_0 ={}  \sqrt{\frac{8\pi G}{3} \, \rho_c}
= 2.1976 \by 10^{-18} \, \text{s}^{-1}  
\end{equation} 
in which $ \rho_c $ 
is the critical density (\ref{critical density}).
The ratio of these frequencies
in $\Lambda$CDM cosmology is
\begin{equation}
\frac{H_{\Lambda\text{CDM}}(a)}
{H_0} ={}
\sqrt{\Omega_{\Lambda} 
+ \Omega_{k} \, a^{-2} 
+ \Omega_{m} \, a^{-3}  
+ \Omega_{r} \, a^{-4} } .
\label {LambdaCDM ratio}
\end{equation} 

\par
In DESI cosmology,
the corresponding ratio of Hubble frequencies is
\begin{equation}
\frac{H_{\text{\tiny{DESI}}}(a)}{H_0} ={}
\sqrt{\Omega_{\Lambda} \, 
 a^{-3(1 + w_0 + w_a)} \,
e^{-3 w_a (1-a)} 
+ \Omega_{k} \, a^{-2}   
+ \Omega_{m} \, a^{-3}  
+ \Omega_{r} \, a^{-4} } .
\end{equation}
The ratios for the Pantheon+, Union3 and DESY5 
combinations 
(\ref{Pantheon+}, \ref{Union3}, \ref{DESY5})
are 
\begin{align}
\frac{H_{\text{\tiny{P+}}}(a)}{H_0} ={}&
\sqrt{\Omega_{\Lambda} \, a^{1.374} \,
e^{1.86 (1-a)}  
+ \Omega_{k} \, a^{-2}   
+ \Omega_{m} \, a^{-3}  
+ \Omega_{r} \, a^{-4} } 
\\
\frac{H_{\text{\tiny{U3}}}(a)}{H_0} ={}&
\sqrt{\Omega_{\Lambda} \, a^{2.271} \,
e^{3.27 (1-a)}  
+ \Omega_{k} \, a^{-2}   
+ \Omega_{m} \, a^{-3}  
+ \Omega_{r} \, a^{-4} } 
\\
\frac{H_{\text{\tiny{D5}}}(a)}{H_0} ={}&
\sqrt{\Omega_{\Lambda} \, a^{1.836} \,
e^{2.58 (1-a)}  
+ \Omega_{k} \, a^{-2}   
+ \Omega_{m} \, a^{-3}  
+ \Omega_{r} \, a^{-4} } .
\label {DESY5 ratio}
\end{align} 

\par
The $\Omega$'s of equations
(\ref{LambdaCDM ratio}--\ref{DESY5 ratio})
are given in Table~\ref{Planck table}. 
They are those of the Planck 
collaboration~\citep{AAplanck2024} 
except that neutrinos 
are counted as radiation
$\Omega_r$ since
$c^2 \sum m_\nu < 0.11$ eV
(equation~43 
of~\citep{AAplanck2024}).

\begin{figure}[h!] 
\begin{center}
\textbf{The DESI scale factors closely track $\bos{\Lambda}$CDM}\\
\includegraphics[width=5.0in, trim={0.in 0.in 0in 0.0in}, clip]{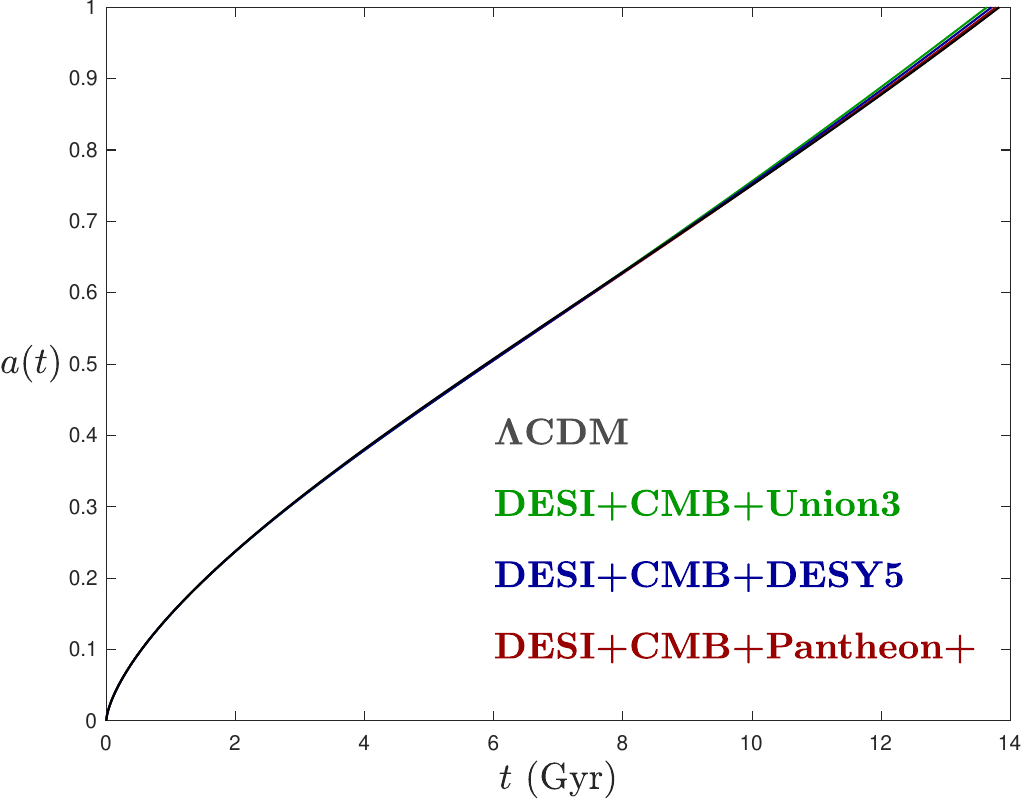}
\caption{For all three choices
\tcfg{DESI+CMB+Union3},
\tcdb{DESI+CMB+DESY5}, and
\textcolor{RawSienna}{DESI+CMB+Pantheon+}
of supernova data, the DESI scale factors
closely track the time evolution
of the scale factor of $\Lambda$CDM.}
\label {fig2}
\end{center}
\end{figure}

\begin{table}[h]
\begin{center}
\textbf{Cosmological parameters}\\
\begin{tabular}{|c|c|c|c|}
\toprule
$ \, H_0$ (km/(s\,Mpc)) \,  & 
$ \Omega_{\Lambda} $  
& $ \Omega_{m} $  & $ \Omega_{r} $ 
\\
$67.81 \pm 0.38$ & $ \,\, 0.6889 \pm 0.0056 \,\, $ & 
$ \,\, 0.3071 \pm 0.0051 \,\, $
& $ \,\, 9.0824 \times 10^{-5} \,\, $
 \\ \bottomrule
\end{tabular}
\end{center}
\caption{Cosmological parameters 
of the $\Lambda$CDM universe
with $\Omega_k = 0.0000 \pm 0.0016$
and with neutrinos considered as radiation 
and included in 
$\Omega_r$~\citep{AAplanck2024}.}
\label {Planck table}
\end{table} 

Although the dark-energy densities
$\rho_{de, \tiny{\Lambda}}(a)$,
$\rho_{de, \tiny{P+}}(a)$, 
$\rho_{de, \tiny{U3}}(a)$, and
$\rho_{de, \tiny{D5}}(a)$ 
of Fig.~\ref{fig1} are very different functions 
of the scale factor $a$, 
their scale factors
$a_{de, \tiny{\Lambda}}(t)$,
$a_{de, \tiny{P+}}(t)$, 
$a_{de, \tiny{U3}}(t)$, and
$a_{de, \tiny{D5}}(t)$
are nearly the same function of 
the time $t$
as illustrated in Fig.~\ref{fig2}.

\begin{figure}[h!]  
\begin{center}
\textbf{The last Gyr}\\
\includegraphics[width=5.0in, trim={0.in 0.in 0in 0.0in}, clip]{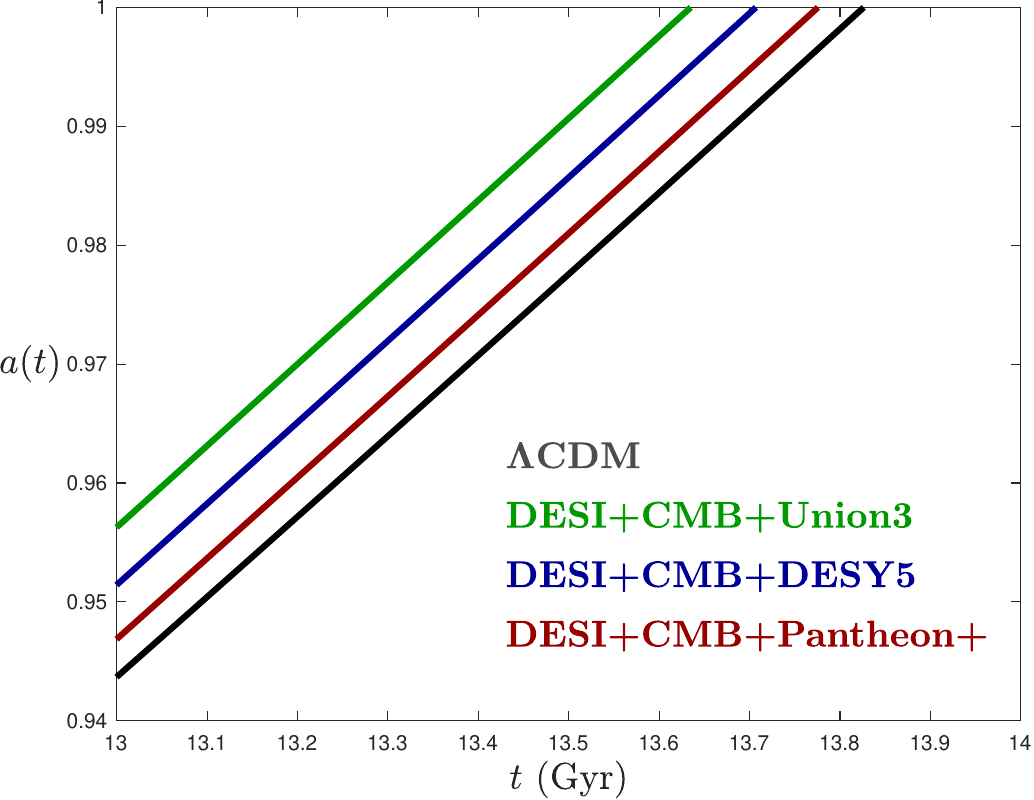}
\caption{The last Gyr 
of the time evolution
of the scale factor $a(t)$
is plotted 
for the $\Lambda$CDM and DESI
cosmologies.
The DESY5 universe is  119 Myr younger
than a $\Lambda$CDM universe.
}
\label {fig3}
\end{center}
\end{figure}

\par

\section{The age of a DESI universe is $\mathbf{13.7 \pm 0.07}$ Gyr}
\label{Two ages for the universe}

\par
In $\Lambda$CDM cosmology,
the age of the universe 
is~\cite{Weinberg2010p34-45, CahillCUP2ageofuniverse}.
\begin{equation}
t_{\Lambda\text{CDM}}(1) ={}  \frac{1}{H_0} \int_0^1
\!\! \frac{da}{ \sqrt{\Omega_{\Lambda} \, a^2
+ \Omega_{k}   
+ \Omega_{m} \, a^{-1}  
+ \Omega_{r} \, a^{-2} } }  
= 13.8 \,\,\text{Gyr} .
\label {LCDM time}
\end{equation}

\par

In DESI cosmology,
the Pantheon+, Union3 
and DESY5 universes are younger 
by 51, 191, and 119 Myr respectively:
\begin{align}
t_{\text{P+}}(1) ={}  \frac{1}{H_0} 
\! \int_0^1
\!\! \frac{da}
{ \sqrt{\Omega_{\Lambda} \, a^{3.374} \,
e^{1.86 (1-a)}  
+ \Omega_{k}   
+ \Omega_{m} \, a^{-1}  
+ \Omega_{r} \, a^{-2} } } 
= 13.776 \,\,\text{Gyr} 
\\
t_{\text{U3}}(1) ={}  \frac{1}{H_0} 
\! \int_0^1
\!\! \frac{da}
{ \sqrt{\Omega_{\Lambda} \, a^{4.271} \,
e^{3.27 (1-a)}  
+ \Omega_{k}   
+ \Omega_{m} \, a^{-1}  
+ \Omega_{r} \, a^{-2} } } 
= 13.635 \,\,\text{Gyr} 
\\
t_{\text{D5}}(1) ={}  \frac{1}{H_0} 
\! \int_0^1
\!\! \frac{da}
{ \sqrt{\Omega_{\Lambda} \, a^{3.836} \,
e^{2.58(1-a)}  
+ \Omega_{k}   
+ \Omega_{m} \, a^{-1}  
+ \Omega_{r} \, a^{-2} } } 
= 13.707 \,\,\text{Gyr} 
\label {DESI time}
\end{align}
as illustrated 
in Fig.~\ref{fig3}.

The Fortran90 codes 
(Planck.f90, 
Pantheon.f90, Union3.f90, and DESY5.f90) 
that compute
the age $t(a)$ of the universe
as a function of the scale factor $a$ 
are available in the DESI 
repository of  
github.com/kevinecahill.

\acknowledgments
I am grateful to Kyle Dawson 
for a useful conversation 
and to Daniel Eisenstein 
for pointing out a misplaced parenthesis
in one of my Fortran90 programs.

\bibliography{physics,math}

\end{document}